\documentclass[english]{article}
\usepackage[T1]{fontenc}
\usepackage{amssymb}
\usepackage{graphicx}

\makeatletter
\usepackage{a4}
\input{epsfig.sty}
\def\fnum@table{\tablename~{\bf\thetable}}
\def\fnum@figure{\figurename~{\bf\thefigure}}
\def\tablename{\footnotesize{\bf Table}}
\def\figurename{\footnotesize{\bf Figure}}

\def\be{\begin{equation}}
\def\ee{\end{equation}}

\makeatother

\usepackage{babel}
\begin{document}

\title{\textbf{On the model uncertainties for the predicted muon content of extensive air showers}}

\author{Sergey Ostapchenko  and G\"unter Sigl\\
\textit{\small Universit\"at Hamburg, II Institut f\"ur Theoretische
Physik, 22761 Hamburg, Germany}\\
}

\maketitle
\begin{center}
\textbf{Abstract}
\par\end{center}
Motivated by the excess of the muon content of cosmic ray induced
 extensive air showers (EAS), relative to EAS modeling,  
observed by the Pierre Auger Observatory,
and by the tension between Auger data and
 air shower simulations on the maximal muon  production depth
   $X^{\mu}_{\max}$, we investigate the possibility  to 
   modify the corresponding EAS simulation results,
 within the Standard Model of particle physics. We start by specifying the 
 kinematic range for secondary hadron production, which is of relevance 
 for such predictions. We further investigate the impact
 on the predicted EAS muon number and on  $X^{\mu}_{\max}$
 of various modifications of the treatment of hadronic interactions,
 in the framework of the QGSJET-III model, in particular the model calibration
 to accelerator data, the amount of the ``glue'' in the pion, and the
 energy dependence of the pion exchange process. None of the considered
 modifications of the model allowed us to enhance the EAS muon content by
 more than 10\%. On the other hand, for the maximal muon  production depth,
 some of the studied  modifications of particle production
  give rise up to  $\sim 10$ g/cm$^2$ larger  $X^{\mu}_{\max}$ values, which increases the difference
 with Auger observations.

\section{Introduction\label{intro.sec}}
Experimental studies of very high energy cosmic rays (CRs) are traditionally performed
using   extensive air shower (EAS) techniques: measuring various characteristics of
nuclear-electromagnetic cascades initiated by interactions of primary CR particles in the
atmosphere \cite{nag00}. Such investigations necessarily involve an extensive use of numerical simulation tools designed to describe the EAS development \cite{eng11}. An accurate reconstruction of the properties of CR primaries depends thus strongly on the precision of  air shower modeling. 
This is especially so for CR composition studies since the relevant 
EAS characteristics are governed by interactions with air nuclei both
of primary CR particles and of secondary hadrons produced in the shower cascade process. 
Consequently,  a successful determination of the primary 
CR composition is only possible if such hadronic
interactions are correctly described by the corresponding Monte Carlo (MC) generators.

One of the traditional methods for CR composition studies relies on measurements of the
muon component of air showers by ground-based detectors. However, for   ultra-high
energy cosmic rays (UHECRs), the use of that technique remained hampered for nearly two
decades by the persistent contradiction between experimental observations and EAS simulations,
regarding the muon number at observation levels \cite{aab15,aab16,PierreAuger:2024neu,abb18}, the discrepancy
reaching 50\% level. Using present air shower simulation tools, the observed muon number $N_{\mu}$ would correspond to primary particles being as heavy
as uranium nuclei, which, given the exceedingly rare natural abundance of such nuclei, appears very unlikely.
A number of theoretical approaches potentially capable to resolve this
so-called ``muon puzzle'' \cite{alb22} has been proposed, typically advocating to physics beyond the Standard Model \cite{far13,anc17,pie21,man23}. 
However, since such suggestions are of a rather speculative character,
it may be important to perform a quantitative study of the uncertainty range for
the predicted  EAS muon content, within the standard physics picture.

Such an investigation is the subject of the current work. Using the recently developed
QGSJET-III model \cite{ost24,ost24a}, we study the dependence of the
 predicted $N_{\mu}$ on uncertainties
regarding the model calibration to relevant accelerator data and
 on possible changes of the underlying theoretical mechanisms. In addition to
the total muon number at ground level, we consider also the maximal muon production
depth $X^{\mu}_{\max}$ measured by the Pierre Auger Observatory \cite{aab14,Collica:2016bck},
for which significant discrepancies with  predictions of EAS
 simulations have also been observed. 
 Compared to other 
 studies of uncertainties of EAS predictions (e.g.\ \cite{ulr11,rie24}),
 our investigation differs in
three key aspects: i) the changes of the corresponding modeling are performed at a 
microscopic level; ii) the considered modifications are restricted by the requirement
not to contradict basic physics principles; iii) the consequences of such changes,
regarding a potential (dis)agreement with relevant accelerator data, are analyzed.

The paper is organized as follows. In Section \ref{kinem.sec},
 we study the kinematic range for secondary
hadron production, which is of relevance to $N_{\mu}$ and $X^{\mu}_{\max}$
predictions, using the QGSJET-II-04 \cite{ost11,ost13}, EPOS-LHC \cite{pie15}, and SIBYLL-2.3
\cite{rie20} MC generators. In Section \ref{varia.sec}, 
we consider potential variations  both of the
parameters and of the underlying theoretical mechanisms of the QGSJET-III model, while checking
the corresponding impact on the predicted  $N_{\mu}$ and $X^{\mu}_{\max}$. Finally, we
discuss our results and conclude in Section \ref{summary.sec}.

 \section{Kinematics of secondary hadron production, relevant to $N_{\mu}$ 
 and $X^{\mu}_{\max}$ predictions\label{kinem.sec}}
The muon component of extensive air showers is formed by a long nuclear cascade in the atmosphere, primarily driven by pion-air interactions.\footnote{Since the multiplicity of secondary kaons in hadron-nucleus and nucleus-nucleus collisions is about 10\% of that  of pions, kaon-air interactions play a somewhat subdominant role.} It is thus evident that the 
multiplicity of secondary pions or, more generally, of all kinds of relatively stable
hadrons (i.e., those whose interaction length in the atmosphere is substantially shorter
than the decay length), has a direct impact on the expected muon signal at ground, which
can be qualitatively understood using the simple Heitler's picture of the cascade process
\cite{hei54} (see also \cite{mat05}). A more delicate question addressed, e.g., in \cite{hil97}
(see also \cite{rei21} for a recent study)
is what range of   energy fractions taken by secondary pions from their parent hadrons,
$x_E=E/E_0$, is of   highest importance for the predicted  $N_{\mu}$. On the one side,
secondary hadron production in high energy collisions is spread over the whole available
rapidity range, $0<y<\ln s$, in the laboratory (lab.) frame, $s$ being the center-of-mass (c.m.) energy squared for the collision. Hence, the bulk of  secondaries is characterized by
vanishingly small $x_E \propto e^{y}/s$. On the other hand, interactions of most energetic secondary hadrons
initiate powerful enough subcascades, thereby providing a higher contribution to the observed
 $N_{\mu}$. The competition between these two trends can be easily quantified considering,
 for the moment, only contributions to muon production from interactions of secondary pions with air and making a rather reasonable
 assumption, being again motivated by Heitler's picture, that the energy-dependence
 of the muon content of a  nuclear subcascade initiated by a charged pion 
 can be approximated by a power law:
 \begin{equation}
N_{\mu}^{\pi^{\pm}}(E_0) \propto E_0^{\alpha_{\mu}}. 
\label{power-spec.eq}
\end{equation}

\begin{figure}[htb]
\centering
\includegraphics[height=6cm,width=0.5\textwidth]{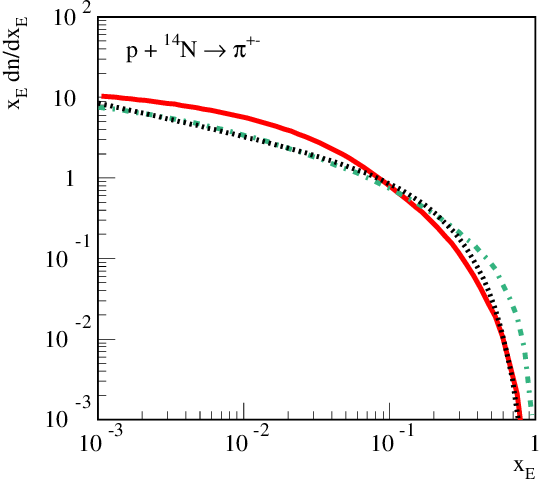}
\caption{$x_E$ distribution of charged pions for $p$N collision,
 $x_E dn^{\pi^{\pm}}_{p{\rm N}}/dx_E$, at $E_0=10^{17}$ eV,
 as calculated using the QGSJET-II-04 (solid line) and SIBYLL-2.3
  (dash-dotted line) models. Shown by the dotted line is the 
  approximation of the SIBYLL-2.3 results, using the ansatz of
   Eq.\ (\ref{xe-distr.eq}),  with  $\Delta = 0.4$ and $\beta = 4.5$.}
\label{fig:xe-fit}      
\end{figure}

 Under these assumptions,   $N_{\mu}$
of a proton-initiated air shower is given as
\begin{equation}
 N_{\mu}^{p}(E_0) \propto \int \!dx_E\; \frac{dn^{\pi^{\pm}}_{p-{\rm air}}(E_0,x_E)}{dx_E}\,
 N_{\mu}^{\pi^{\pm}}(x_E E_0) \,,
 \label{nmu-p.eq}
\end{equation}
with $dn^{\pi^{\pm}}_{p-{\rm air}}(E_0,x_E)/dx_E$ being the  $x_E$ distribution of
secondary charged pions in a proton-air collision at lab.\ energy $E_0$.
Such a distribution, for hadron $h$ collisions with air nuclei, can be  approximated as
\begin{equation}
 \frac{dn^{\pi^{\pm}}_{h-{\rm air}}(E_0,x_E)}{dx_E}\propto x_E^{-1-\Delta_h(E_0)}\,
 (1-x_E)^{\beta_h(E_0)} \,,
 \label{xe-distr.eq}
\end{equation}
within the differences between the predictions of current MC generators,
as illustrated in Fig.\ \ref{fig:xe-fit} 
 for the case of a proton-nitrogen ($p$N) collision at $E_0=10^{17}$ eV.

 Inserting Eq.\ (\ref{xe-distr.eq}) into Eq.\ (\ref{nmu-p.eq})
and making use of Eq.\ (\ref{power-spec.eq}),
we have immediately
\begin{equation}
 N_{\mu}^{p}(E_0) \propto  E_0^{\alpha_{\mu}}\, \int \!dx_E\; x_E^{\alpha_{\mu}-1-\Delta(E_0)}\,
 (1-x_E)^{\beta(E_0)} \,.
 \label{nmu-p-appr.eq}
\end{equation}

\begin{figure}[htb]
\centering
\includegraphics[height=6cm,width=0.5\textwidth]{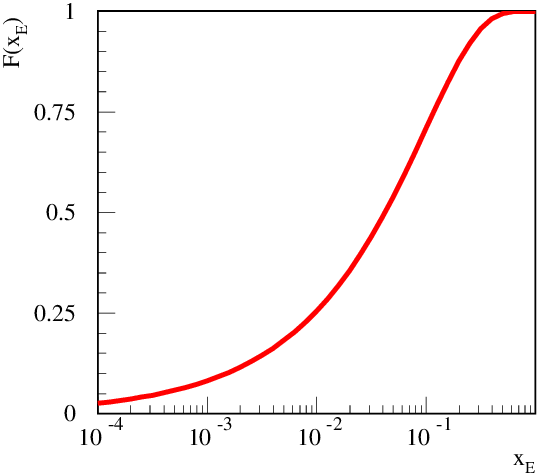}
\caption{Cumulative $x_E$ distribution, Eq.\ (\ref{x-cumul.eq}), 
for  $\Delta = 0.4$, $\beta = 4.5$, and $\alpha_{\mu} = 0.9$.}
\label{fig:xe-cumul}      
\end{figure}

\begin{figure}[htb]
\centering
\includegraphics[height=6cm,width=0.5\textwidth]{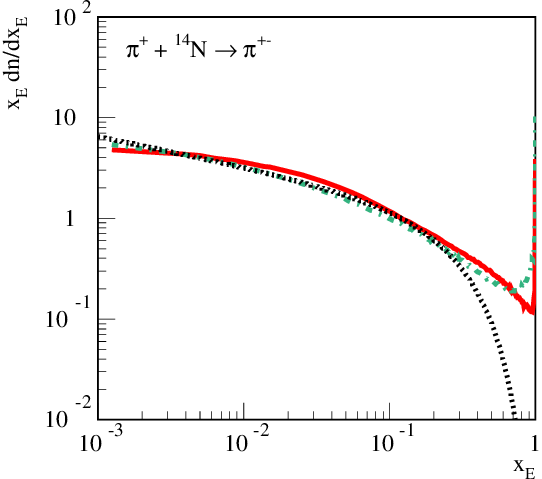}
\caption{$x_E$ distribution of charged pions for $\pi^+$N collision,
 $x_E dn^{\pi^{\pm}}_{\pi^+{\rm N}}/dx_E$, at $E_0=1$ PeV,
 as calculated using the QGSJET-II-04 (solid line) and SIBYLL-2.3
  (dash-dotted line) models. Shown by the dotted line is the 
  approximation of the SIBYLL-2.3 results, using the ansatz of
   Eq.\ (\ref{xe-distr.eq}),  with  $\Delta = 0.3$ and $\beta = 3.5$.}
\label{fig:xe-fit-pi}      
\end{figure}

From Eq.\ (\ref{nmu-p-appr.eq}), we get the ``average'' $\langle x_E^{\pi}\rangle$ for
$p$-air interactions,  weighted by the contributions of the corresponding pion
subcascades to  $N_{\mu}$:
\begin{equation}
 \langle x_E^{\pi}\rangle = \frac{\int_0^1 \!dx_E\; x_E^{\alpha_{\mu}-\Delta}\,
 (1-x_E)^{\beta}}{\int_0^1 \!dx_E\; x_E^{\alpha_{\mu}-1-\Delta}\,
 (1-x_E)^{\beta}}=\frac{\alpha_{\mu}-\Delta}{1+\alpha_{\mu}+\beta-\Delta}\,.
 \label{x-pi-avr.eq}
\end{equation}
For   $\alpha_{\mu} \simeq 0.9$, using the parameters of the fit in 
Fig.\ \ref{fig:xe-fit}, $\Delta = 0.4$ and $\beta = 4.5$,
we get $\langle x_E^{\pi}\rangle \simeq 0.08$.
 In Fig.\ \ref{fig:xe-cumul},
 we plot the cumulative $x_E$ distribution, 
\begin{equation}
F(x_E) = \frac{\int_0^{x_E} \!dx \; x^{\alpha_{\mu}-1-\Delta}\,
 (1-x)^{\beta}}{\int_0^1 \!dx\; x^{\alpha_{\mu}-1-\Delta}\,
 (1-x)^{\beta}}\,,
 \label{x-cumul.eq}
\end{equation}
for these parameter values. As one can see in the Figure, only 1/4 
of the contribution comes from the interval $x_E<0.01$.
The same reasoning can be repeated for
pion-air collisions, if we neglect the diffractive contribution which, 
as will be demonstrated in the following,
 is of minor importance for  $N_{\mu}$, yielding a similarly
  large value for  $\langle x_E^{\pi}\rangle$.
  This is illustrated in  Fig.\ \ref{fig:xe-fit-pi},
 where the $x_E$ distribution of charged pions for $\pi^+$N collisions
 at $E_0=1$ PeV is approximated by the ansatz of
   Eq.\ (\ref{xe-distr.eq}),  with  $\Delta = 0.3$ and $\beta = 3.5$.
   Using these values of  $\Delta$ and $\beta$ in Eq.\ (\ref{x-pi-avr.eq})
   yields  $\langle x_E^{\pi}\rangle \simeq 0.12$.

 Thus, the air shower muon content is governed by forward pion production\footnote{Forward
 production traditionally refers to a creation of secondary particles in the projectile
 fragmentation region, which is characterized by finite $x_E$ values in lab.\ frame.
 In contrast, central production refers to an emission of secondaries at small values
 of rapidity in c.m.\ frame, corresponding to vanishingly small  $x_E$ in the lab.\ frame,
  in the very high energy limit.} in hadron-air
 collisions, at the many steps of the nuclear cascade in the atmosphere. Let us further
 support this conclusion by a numerical study, using a number of popular MC generators
 of CR interactions. To this end, we modify  pion energy distributions 
 provided by
 those models, replacing every second pion characterized by  energy fraction $x_E$
 bigger than some cutoff $x_{\rm cut}$ by a pair of pions of half the energy. Importantly,
 doing so, we replace a charged pion by a pair of charged pions, while a  $\pi^0$
 is being ``split'' into a pair of neutral pions, such that the energy partition between 
 nuclear and electromagnetic (e/m) cascades remains unaltered. 
 
\begin{figure}[htb]
\centering
\includegraphics[height=6cm,width=0.5\textwidth]{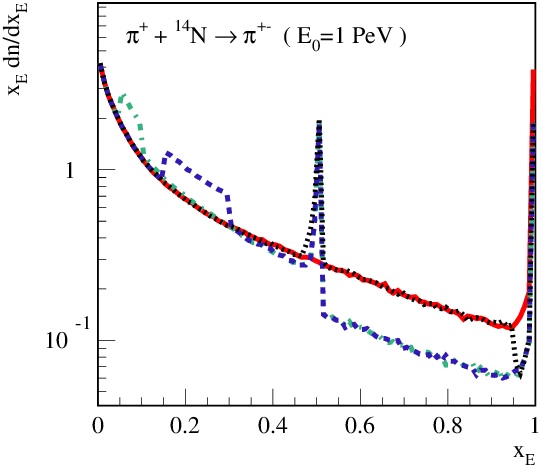}
\caption{$x_E$ distribution of charged pions for $\pi^+$N collision,
 $x_E dn^{\pi^{\pm}}_{\pi{\rm N}}/dx_E$, at $E_0=1$ PeV,
 as calculated using the QGSJET-II-04 model (solid line).
 The modifications of this distribution by the ``splitting'' procedure
 discussed in the text are shown by the dotted, dashed, and dash-dotted lines
 for  $x_{\rm cut}=0.95$, 0.3, and 0.1, respectively.}
\label{fig:split}      
\end{figure}

 The modifications of secondary pion energy distributions by such a procedure
  are exemplified in Fig.\ \ref{fig:split}.
 For large $x_{\rm cut}\simeq 1$, this impacts
 the diffractive peak only, leaving  a half of it;
  such a peak reappears then at $x_E\simeq 0.5$. Choosing a smaller $x_{\rm cut}$, a larger and larger part of the very forward pion spectrum is reduced, being
 accompanied by an enhancement of pion production at smaller $x_E$. In the limit 
  $x_{\rm cut}\rightarrow 0$,   the forward pion production is reduced considerably, while the 
 total multiplicity of secondary pions is increased by 50\%.
 
\begin{figure}[htb]
\centering
\includegraphics[height=6cm,width=0.5\textwidth]{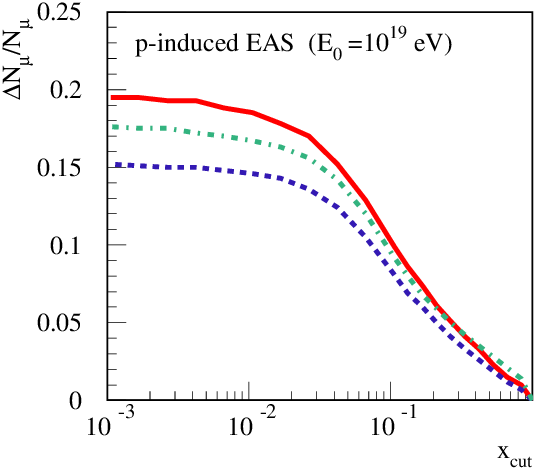}
\caption{$x_{\rm cut}$ dependence of the relative change of $N_{\mu}$ ($E_{\mu}>1$ GeV) at sea level, for $p$-induced  EAS of $E_0=10^{19}$ eV, 
for the pion ``splitting'' procedure, 
for different interaction
models: QGSJET-II-04 (solid line), EPOS-LHC (dashed line), and SIBYLL-2.3 (dash-dotted line).}
\label{fig:xe-nmu}      
\end{figure}

\begin{figure}[htb]
\centering
\includegraphics[height=6cm,width=0.5\textwidth]{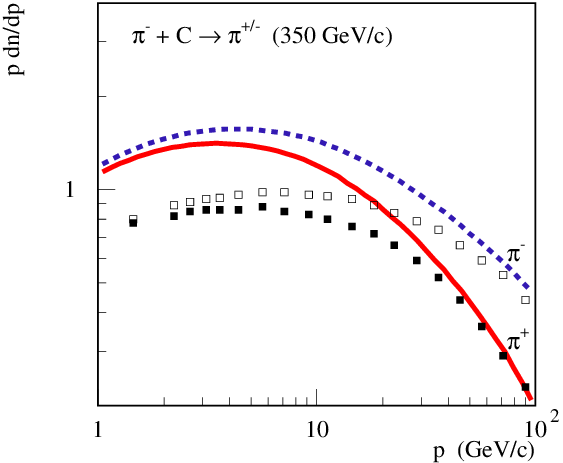}
\caption{Momentum distributions  (in lab.\ frame) of charged pions  produced
  in $\pi^-C$ collisions at  350 GeV/c,   obtained  using the  QGSJET-II-04 model
  and applying the ``splitting'' procedure to  half the secondary pions,
    compared to NA61 data \cite{adh23} (points):
$\pi^+$ --  solid line and filled squares,  $\pi^-$ -- 
   dashed line and open squares.}
\label{fig:modpion}      
\end{figure}

 In Fig.\ \ref{fig:xe-nmu}, 
 we plot the $x_{\rm cut}$ dependence of the relative change
 of the  calculated  $N_{\mu}$ ($E_{\mu}>1$ GeV) at sea level,
  $\Delta N_{\mu}/N_{\mu}$, for $p$-induced EAS of $E_0=10^{19}$ eV, 
   applying such a ``splitting'' procedure to all
 hadron-air interactions in the shower,\footnote{Here and in the following, we perform EAS
 simulations, using the CONEX code \cite{ber07}.} which are treated by 
 the QGSJET-II-04, EPOS-LHC, or SIBYLL-2.3 models. 
 The first thing to notice is that for $x_{\rm cut}$ values close to unity,
 there is practically no change of $N_{\mu}$, i.e., pion diffraction is rather irrelevant
 for EAS muon content. Secondly, 
 the largest gradient
  of the  $\Delta N_{\mu}(x_{\rm cut})$ dependence is 
 observed for $x_{\rm cut}\sim 0.1$,  thereby supporting our
 previous estimation, Eq.\ (\ref{x-pi-avr.eq}). In the limit $x_{\rm cut}\rightarrow 0$, the 
 EAS muon content is enhanced by $\sim 15-20$\%, depending on the interaction model.
  It should be stressed, however, that the
 performed study serves illustrative purposes only, not being a viable procedure for
 modifying the predicted $N_{\mu}$. For example, when the discussed ``splitting'' is
 applied to half the secondary pions (the case  $x_{\rm cut}\rightarrow 0$), one
  arrives to a  strong contradiction with available accelerator data. 
 This is illustrated in Fig.\ \ref{fig:modpion},
 where we plot momentum distributions  of charged pions,
 for $\pi^-C$ collisions at  350 GeV/c,   obtained  using the  QGSJET-II-04 model
  and applying the ``splitting'' procedure to  half the secondary pions (the case
 $x_{\rm cut}=0$), in   comparison to NA61 data \cite{adh23}.

\begin{figure}[htb]
\centering
\includegraphics[height=6cm,width=0.5\textwidth]{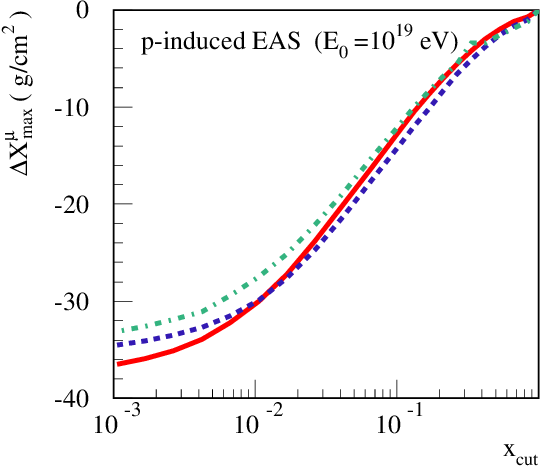}
\caption{$x_{\rm cut}$ dependence for the modification of  $X^{\mu}_{\max}$ 
  by the pion ``splitting'' procedure, 
for $p$-induced EAS of $E_0=10^{19}$ eV, for different interaction models.
 The meaning of the lines is the same as in Fig.\ \ref{fig:xe-nmu}.}
\label{fig:xe-xmumax}       
\end{figure}

The same kind of analysis has been performed for the maximal muon production
depth $X^{\mu}_{\max}$, the obtained $\Delta X^{\mu}_{\max}(x_{\rm cut})$ dependence being
plotted in Fig.\ \ref{fig:xe-xmumax}, 
again for  $p$-induced EAS of $E_0=10^{19}$ eV. Somewhat surprisingly,
for this dependence, the maximal gradient is observed  in almost the same $x_{\rm cut}$ range: of maximal importance for $X^{\mu}_{\max}$  are secondary pions carrying few percent of the parent hadron energy. As discussed in \cite{ost16},  $X^{\mu}_{\max}$ roughly corresponds to the depth in the atmosphere,
where the interaction and decay lengths for most of the pions in the cascade become comparable, i.e., where the bulk of the pions approaches the pion ``critical'' energy. Hence, this quantity is  sensitive both to the pion-air inelastic cross section and to forward
production spectrum of pions. What was not properly realised in  that
work  is that  the position of that ``equilibrium'' point can be changed most efficiently
by modifying how fast the energy partition between the produced secondary hadrons is established,
thereby speeding up the decrease of the average pion energy, with increasing depth,
which can again be understood using the simple Heitler's picture for a nuclear cascade.
Here, similarly to the case of $N_{\mu}$, one has a competition between a copious hadron
production for $x_E\rightarrow 0$ and much less numerous secondaries characterised by finite
$x_E$, which, however, have a much stronger impact on the speed of 
decrease of the average pion energy.
It is this competition which leads to the obtained range of pion
energy fractions, corresponding to the maximal relevance for  $X^{\mu}_{\max}$ predictions.
As a side remark, pion diffraction appears to have practically no impact on the predicted
$X^{\mu}_{\max}$, contrary to what had been claimed in the literature \cite{pie17}
(cf.\ Fig.\  \ref{fig:xe-xmumax} for $x_{\rm cut}\rightarrow 1$).

\begin{figure}[htb]
\centering
\includegraphics[height=6cm,width=0.5\textwidth]{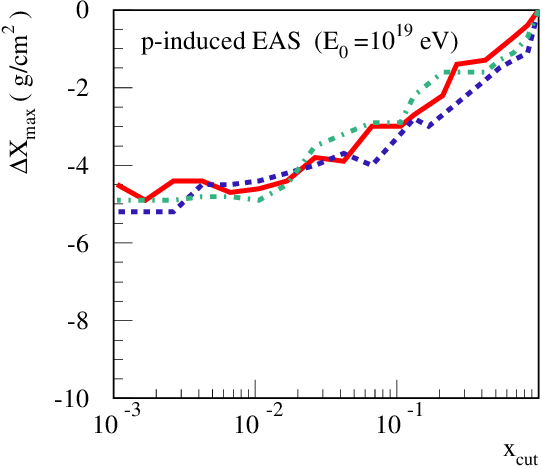}
\caption{$x_{\rm cut}$ dependence for the modification of  $X_{\max}$ 
   by the pion ``splitting'' procedure, 
for $p$-induced EAS of $E_0=10^{19}$ eV, for different interaction models.
 The meaning of the lines is the same as in Fig.\ \ref{fig:xe-nmu}.}
\label{fig:xe-xmax}       
\end{figure}

For completeness, we plot in Fig.\ \ref{fig:xe-xmax} 
the $x_{\rm cut}$ dependence for the change of the calculated
 EAS maximum depth $X_{\max}$, under  such a  ``splitting'' procedure. 
 As one can see in the Figure, the considered modifications of secondary pion
energy spectrum have a very weak impact on the predicted $X_{\max}$, i.e., the treatment
of pion-air interactions is of small importance for $X_{\max}$ predictions:
even in the extreme case,  $x_{\rm cut}\rightarrow 0$, the  EAS maximum depth changes by
less than 5 g/cm$^2$.

To summarize this part of our analysis, both $N_{\mu}$ and   $X^{\mu}_{\max}$
 are governed by forward pion production in hadron-air collisions:
  with the corresponding pion energy fractions ranging between few
   per cent and few tens per cent. Obviously, these conclusions can be generalized by taking into consideration
the production of all ``stable'' hadrons, i.e., the obtained ranges of
energy fractions $x_E$ relevant for $N_{\mu}$ and   $X^{\mu}_{\max}$ predictions apply to
the production spectra of all such hadrons.

\begin{figure}[htb]
\centering
\includegraphics[height=6.cm,width=\textwidth]{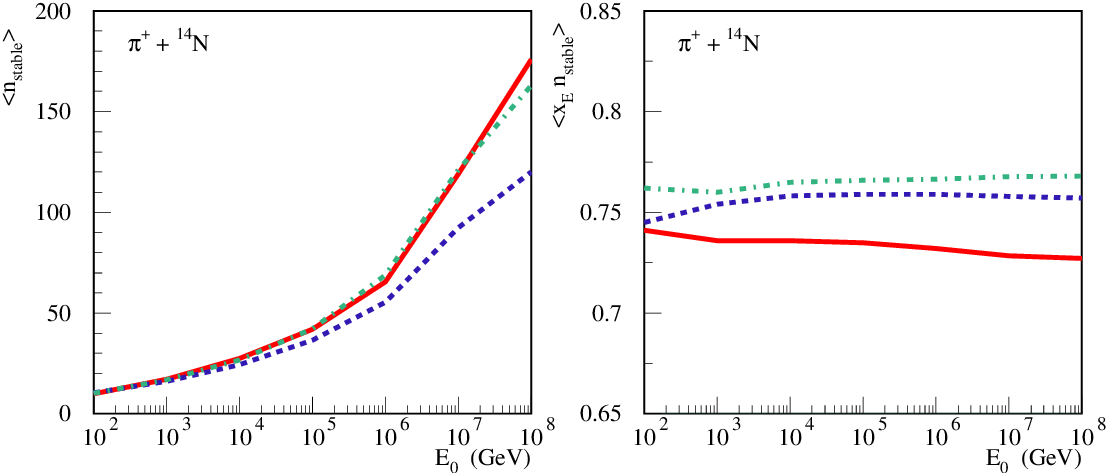}
\caption{Energy-dependence of the multiplicity of ``stable'' hadrons (left)
 and of the energy fraction taken by all 
 such  hadrons (right) for $\pi^+$N
  interactions,  calculated using the QGSJET-II-04, EPOS-LHC, and 
  SIBYLL-2.3 models -- solid, dashed, and dash-dotted lines, respectively.}
\label{fig:nstable}       
\end{figure}

\begin{figure}[t]
\centering
\includegraphics[height=6cm,width=0.5\textwidth]{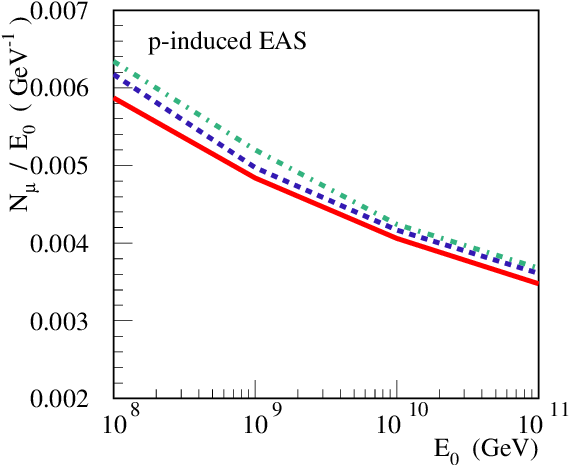}
\caption{Dependence on primary energy of the muon number  $N_{\mu}$
 (at sea level) of proton-initiated EAS, for $E_{\mu}>1$ GeV,
  calculated using  the QGSJET-II-04, EPOS-LHC, and SIBYLL-2.3 
  models -- solid, dashed, and dash-dotted lines, respectively.}
\label{fig:nmu}       
\end{figure}

To further support these conclusions, we plot in Fig.\ \ref{fig:nstable} the 
energy-dependence of the multiplicity of ``stable'' hadrons 
[(anti)nucleons, kaons, and charged pions]
$\langle n_{\rm stable}^{\pi N}\rangle$  and of the second moment of the corresponding 
energy fraction distribution,\footnote{In view of some (weak)  energy and model dependence
of the exponent $\alpha_{\mu}$ [cf.\ Eq.\ (\ref{power-spec.eq})], we choose to use
$\langle x_E\, n_{\rm stable}^{\pi N}\rangle$, instead of 
$\langle x_E^{\alpha_{\mu}} n_{\rm stable}^{\pi N}\rangle$ 
considered in  \cite{hil97,rei21}.}
 $dn_{\rm stable}^{\pi N}/dx_E$, i.e., the average 
energy fraction taken by all stable secondary hadrons, for pion-nitrogen collisions, 
\begin{equation}
 \langle x_E\, n_{\rm stable}^{\pi N}(E_0)\rangle =  \int \!dx_E\; x_E\,
 \frac{dn_{\rm stable}^{\pi N}(E_0,x_E)}{dx_E}\,, 
 \label{x*N.eq}
\end{equation}
 calculated using the QGSJET-II-04, EPOS-LHC, and SIBYLL-2.3 models. 
As one can see in the Figure, the most steep energy rise of $\langle n_{\rm stable}^{pN}\rangle$ is observed for
  QGSJET-II-04, while EPOS-LHC and SIBYLL-2.3 predict noticeably higher
values of $\langle x_E\, n_{\rm stable}^{\pi N}\rangle$ -- due to a much more
abundant forward production of, respectively, (anti)nucleons and
 $\rho$ mesons in those models.\footnote{See, however, the
corresponding criticism in \cite{ost23}.} The more abundant forward  production 
of stable hadrons
 in the latter two models clearly explains their somewhat higher values of the predicted 
  $N_{\mu}$,   plotted in  Fig.\ \ref{fig:nmu}.
Thus,  it is the total fraction of the incident hadron energy, taken by all stable secondary hadrons, rather than the multiplicity, which governs the muon content of extensive air showers.
  
 \section{Potential variations of model predictions for $N_{\mu}$ and   $X^{\mu}_{\max}$\label{varia.sec}}
 \subsection{Enhancing the gluon content of the pion\label{pion-glue.sec}}
As is evident from the analysis in Section \ref{kinem.sec}, to increase the
 predicted  $N_{\mu}$, one has to enhance forward production of stable 
 hadrons. Generally, there exists
no viable theoretical mechanism to produce a ``hardening'' of secondary 
hadron spectra, with increasing energy: 
the energy-rise of   multiple scattering   should rather give rise
 to a (rather moderate) opposite
effect, leading to the energy partition between larger and larger numbers
 of secondaries. Therefore, to obtain an enhancement of forward
  production, one may try to increase the
multiple scattering rate in pion-air collisions, 
desirably, without a significant modification of
 the shape of  forward spectra: 
 in order to have higher secondary particle yields
 for all values of $x_E$. 
 Since the energy-rise of multiple scattering is primarily
 driven by (semi)hard production processes giving rise to an emission of hadron (mini)jets,
 the only efficient and, probably, the only possible way to reach such a goal, without
 modifying substantially the treatment of proton-air collisions, rather seriously constrained
 by available data from the Large Hadron Collider (LHC), is to enhance the gluon content
 of the pion.
 
 In principle, the fraction of the pion momentum, carried by its valence 
 quarks,  $\langle x_{q_v}\rangle$, is already
 strongly constrained (e.g.\ \cite{glu99}), notably, by experimental studies of the 
 Drell-Yan and direct photon production processes in pion-proton interactions.
 Therefore, one may only change the partition of the remaining
 momentum fraction, $1-\langle x_{q_v}\rangle$, between gluons and sea (anti)quarks,
 since experimental data on pion structure functions are rather scarce and correspond
 to relatively large values of Bjorken $x$. Shifting that momentum balance towards gluons,
 at the expense of sea (anti)quarks, would enhance the multiple scattering rate: because
 gluons participate in hard scattering with a larger color factor, 
 $C_A=3$, compared to  $C_F=4/3$ of (anti)quarks.
 
\begin{figure}[htb]
\centering
\includegraphics[height=6cm,width=0.5\textwidth]{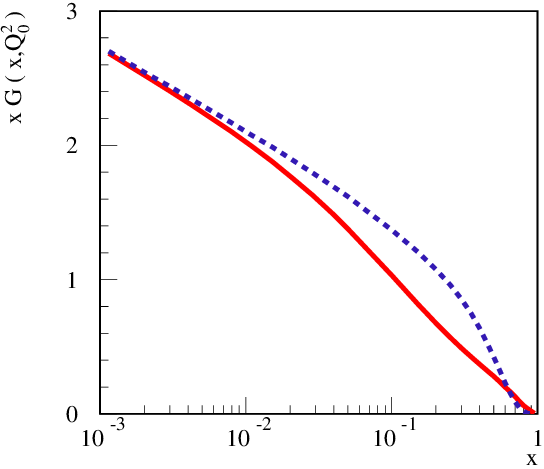}
\caption{Gluon PDF of the pion at $Q_0^2=2$ GeV$^2$ in the default QGSJET-III model
(solid line) and for the enhanced gluon content of the pion,
as discussed in the text (dashed line).}
\label{fig:glupdf}       
   \end{figure}

 Here we wish to study an extreme variation of the glue in the pion. Therefore, we choose
 to modify the overall momentum fraction possessed by both gluons and  sea (anti)quarks,
 $\langle x_{g}\rangle + \langle x_{q_s}\rangle$, at the expense of $\langle x_{q_v}\rangle$.
 I.e., for the moment, we disregard the respective experimental constraints and consider
 a factor of two reduction of $\langle x_{q_v}\rangle$,\footnote{This is achieved by
 modifying the original GRS parton distribution functions (PDFs) of valence quarks
 in the pion \cite{glu99}, employed in the  QGSJET-III model,
  at the cutoff scale $Q_0^2=2$ GeV$^2$ for hard processes,
 by a factor $A_{\delta}(1-x)^{\delta}$, and choosing the parameters $\delta$ and  
 $A_{\delta}$ in such a way that the desirable reduction of  $\langle x_{q_v}\rangle$
 is obtained without violating the valence quark sum rule. Additionally, we increase
 the value of the parameter $\beta_{\pi}$ governing the ``hardness'' of the gluon PDF
 in  QGSJET-III ($\propto (1-x)^{\beta_{\pi}}$ -- cf.\ Eq.\ (8) in \cite{ost24}),
 from the default value 2 to 5, in order to shift the enhancement of the glue
 towards smaller $x$.  PDFs at higher virtuality scales
 $q^2$ are  obtained via a DGLAP evolution from $Q_0^2$ to $q^2$.} 
 while enhancing correspondingly
 both $\langle x_{g}\rangle$ and  $\langle x_{q_s}\rangle$.
 This way we change  the gluon PDF  
rather radically in the $x$-range relevant for $N_{\mu}$ and   $X^{\mu}_{\max}$
predictions, as demonstrated  in Fig.\ \ref{fig:glupdf}.

\begin{figure}[htb]
\centering
\includegraphics[height=6.cm,width=\textwidth]{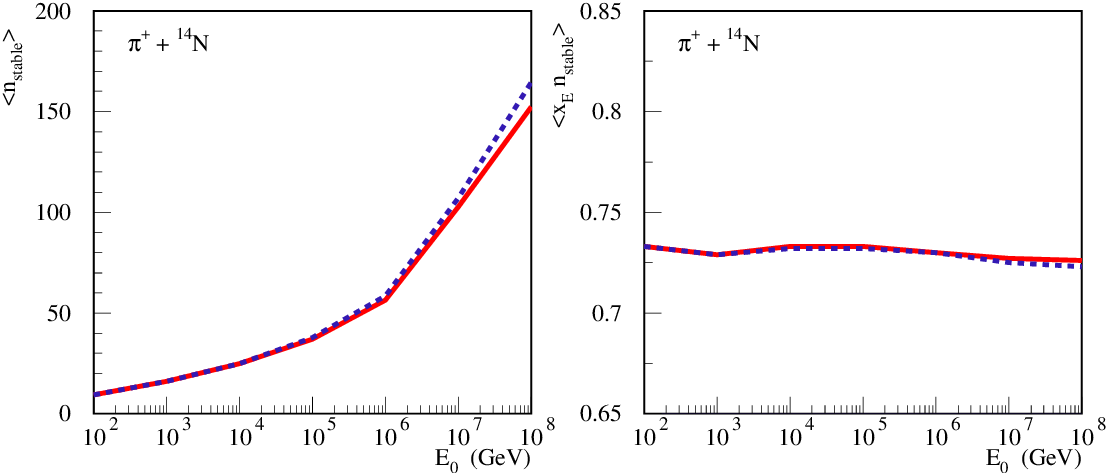}
\caption{Energy-dependence of the multiplicity of ``stable'' hadrons (left)
 and of the energy fraction taken by all 
 such  hadrons (right) for $\pi^+$N
  interactions, for the default QGSJET-III model (solid line) and 
   for the enhanced gluon content of the pion,
as discussed in the text  (dashed line).}
\label{fig:nstableq}       
\end{figure}

Yet, as one can   see in  Fig.\ \ref{fig:nstableq} (left), such a
 modification gives rise to a noticeably
 higher multiplicity of stable secondary hadrons
 $\langle n_{\rm stable}^{\pi N}\rangle$, for pion-nitrogen collisions,
  only at the highest energies considered.
 Moreover, the average fraction of the parent pion energy, taken by all such hadrons,
 $\langle x_E\, n_{\rm stable}^{\pi N}\rangle$, remains practically unaffected,
 see   Fig.\ \ref{fig:nstableq} (right),
 which means that the considered modification has a
minor impact on forward hadron spectra. 
Not surprisingly,  applying the so-modified treatment of pion-air 
interactions to EAS modeling,  we observed rather insignificant
 changes for the calculated EAS muon content and for the
 maximal muon production depth, compared to  QGSJET-III predictions
 ($\lesssim 1$\% for  $N_{\mu}$ and few  g/cm$^2$ for $X^{\mu}_{\max}$).
  
 \subsection{Modifying the model calibration to experimental data\label{model-tune.sec}}
\begin{figure}[htb]
\centering
\includegraphics[height=12cm,width=\textwidth]{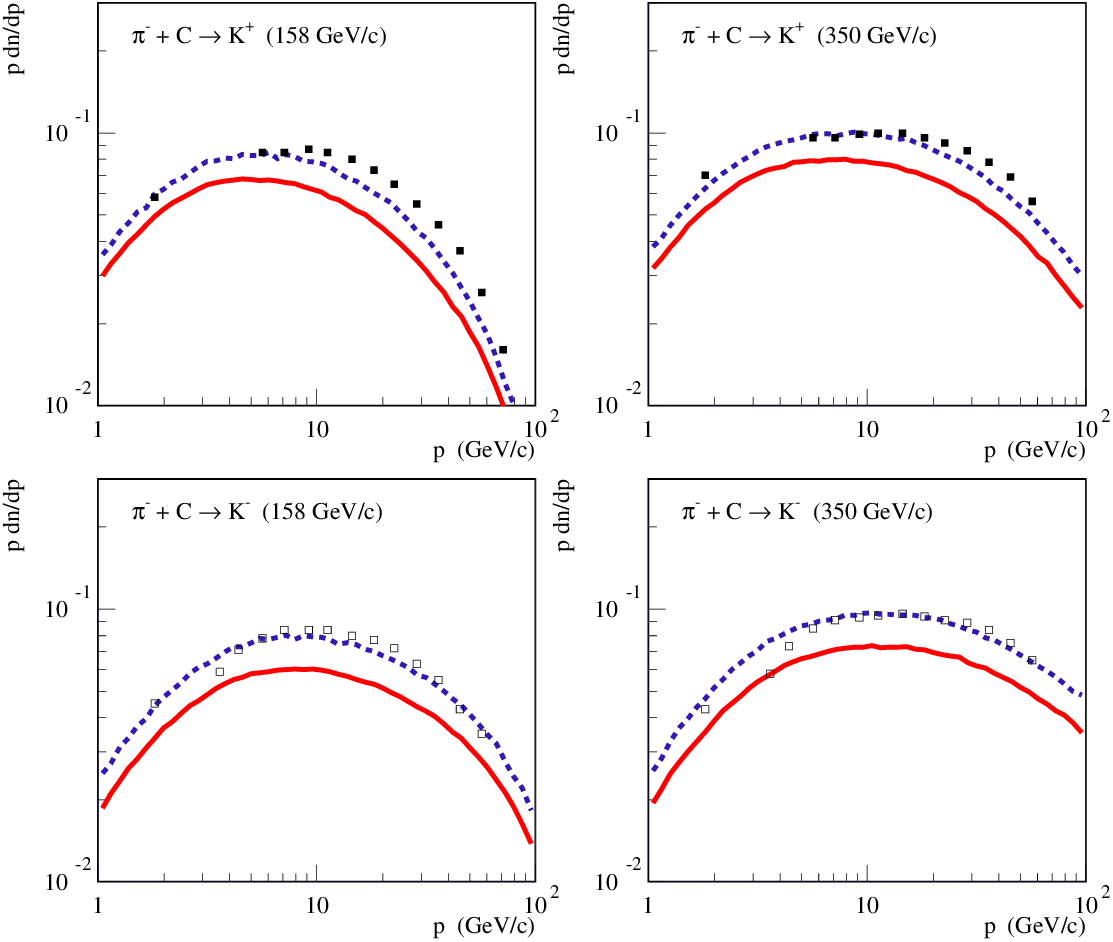}
\caption{Momentum distributions (in lab.\ frame) of $K^+$ (top panels) and  $K^-$ (bottom panels) produced in $\pi^-$C collisions at 158 GeV/c (left) and 350 GeV/c (right),
 as calculated using the default QGSJET-III model
(solid lines) or considering an enhancement of kaon production, as discussed in the text (dashed lines), compared to NA61 data \cite{adh23} (points).}
\label{fig:modk}       
\end{figure}

\begin{figure}[htb]
\centering
\includegraphics[height=12cm,width=\textwidth]{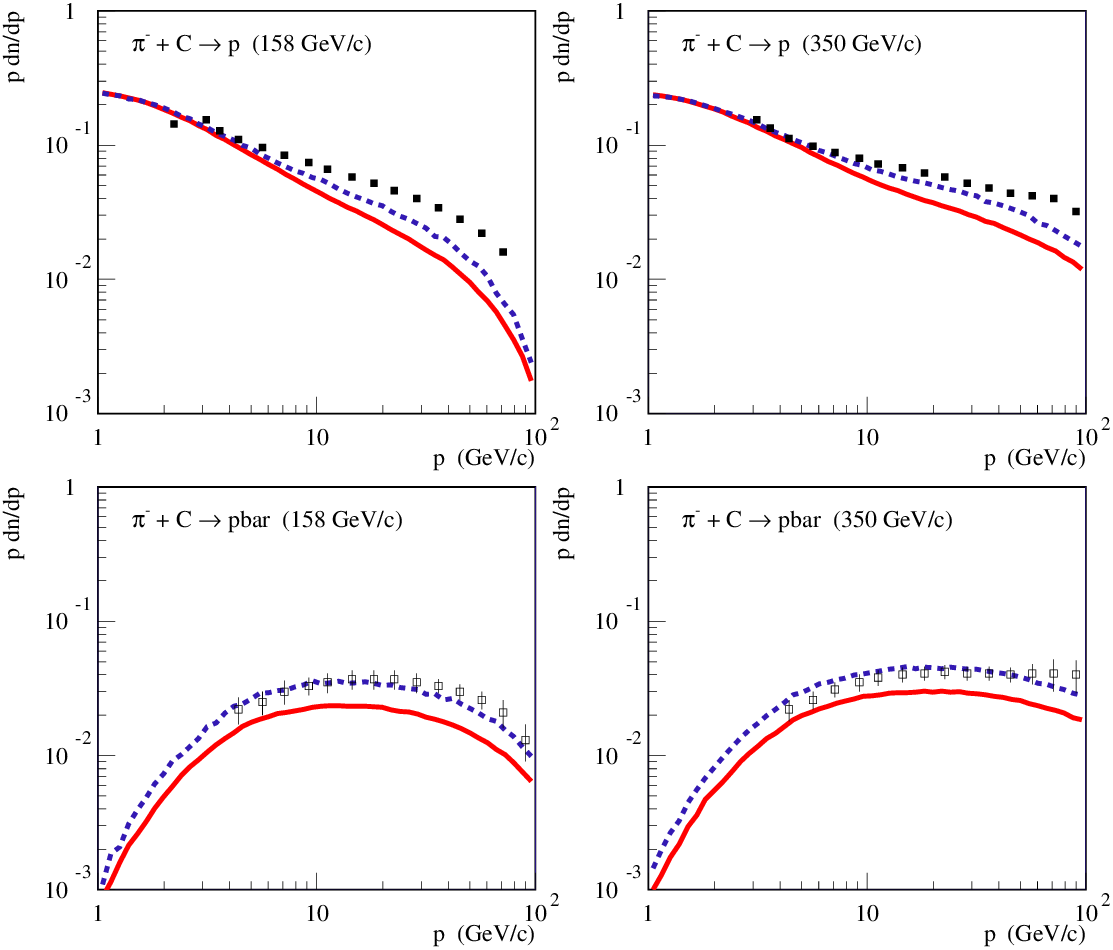}
\caption{Momentum distributions (in lab.\ frame) of protons (top panels) and  antiprotons (bottom panels) produced in $\pi^-$C collisions at 158 GeV/c (left) and 350 GeV/c (right),
 as calculated using the default QGSJET-III model
(solid lines) or considering an enhancement of (anti)nucleon production, 
as discussed in the text (dashed lines), compared to NA61 data \cite{adh23} (points).}
\label{fig:modp}       
\end{figure}

\begin{figure}[htb]
\centering
\includegraphics[height=6cm,width=\textwidth]{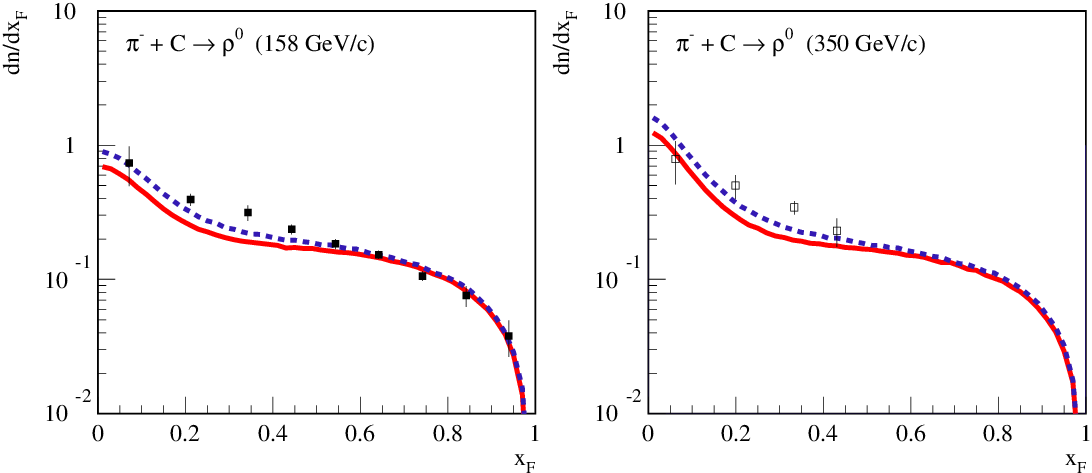}
\caption{Feynman $x$ distributions (in c.m.\ frame) of $\rho^0$ mesons produced in $\pi^-$C
collisions at 158 GeV/c (left) and 350 GeV/c (right), as calculated using the default QGSJET-III model (solid lines) or considering an enhancement
 of $\rho$ meson production, as discussed in the text (dashed lines), 
 compared to NA61 data \cite{adu17} (points).}
\label{fig:modrho}       
\end{figure}

 An alternative way to increase the predicted $N_{\mu}$ is to enhance the
yields of stable secondary hadrons, at the expense of neutral pions. 
In relation to that, let us pay attention to the fact that the  
QGSJET-III  model seriously underestimates the 
 production of kaons and (anti)nucleons in $\pi^-$C collisions,
  compared to measurements by the  NA61 experiment  \cite{adh23}, 
  as one can see in  Figs.\ \ref{fig:modk} and  \ref{fig:modp} (solid lines).
   The agreement with the data can be improved  performing 
   suitable modifications of the
   hadronization procedure of the model:  considering, respectively,
    40\% increase for the
   probability to create strange quark-antiquark ($s\bar s$) pairs from
    the vacuum or 60\%
   enhancement of the corresponding probability for diquark-antidiquark
    ($ud\bar u\bar d$)  pairs, when treating the hadronization of strings of 
    color field, stretched between final partons    produced \cite{ost24a}, 
    the corresponding results being shown in  Figs.\ \ref{fig:modk} and
      \ref{fig:modp} by dashed lines.
   On the other hand, the  agreement with  NA61 data on
     $\rho^0$ meson production \cite{adu17} can be further improved considering 50\%
     enhancement for the probability of the fragmentation of light
      (anti)quarks into $\rho$ mesons, compared to the default value 1/3
      used in the model, see  Fig.\    \ref{fig:modrho}.

\begin{figure}[htb]
\centering
\includegraphics[height=6.cm,width=\textwidth]{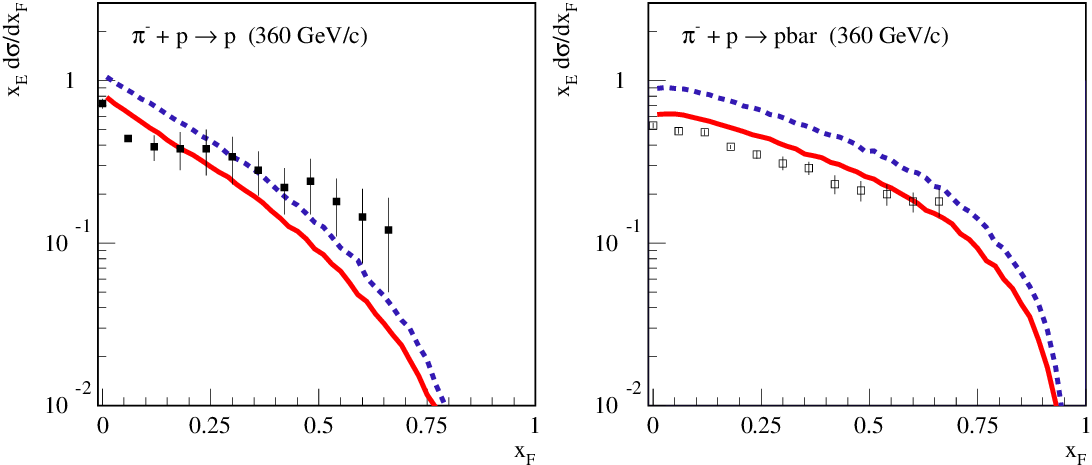}
\caption{Feynman $x$ distributions (in c.m.\ frame) of protons (left) and antiprotons (right)
  in $\pi^-p$ collisions at  360 GeV/c,  as calculated using the default QGSJET-III model
(solid lines) or considering an enhancement of (anti)nucleon production, 
as discussed in the text (dashed lines), compared to LEBC-EHS data \cite{agu87} (points).}
\label{fig:modp360}       
\end{figure}

 It is noteworthy, however, that the considered enhancements of kaon and 
 (anti)nuc\-leon production  are at  tension with the corresponding data on  pion-proton 
 collisions from other experiments, as one can see, e.g., in Fig.\  
 \ref{fig:modp360}. 
 Moreover, since the hadronization procedure is universal 
 for all kinds of hadronic interactions, such changes lead
 to a strong contradiction with the corresponding particle yields 
 measured in $pp$ interactions, both at fixed target energies and at 
 LHC (see \cite{ost24a} for a comparison of the default
 QGSJET-III results with such data). In turn, for the 
 considered enhancement of $\rho$ meson production, the fraction of directly
 produced   pions falls down to $\simeq 30$\%, in a clear contradiction to  observations \cite{agu89}.
 
\begin{figure}[htb]
\centering
\includegraphics[height=6cm,width=0.49\textwidth]{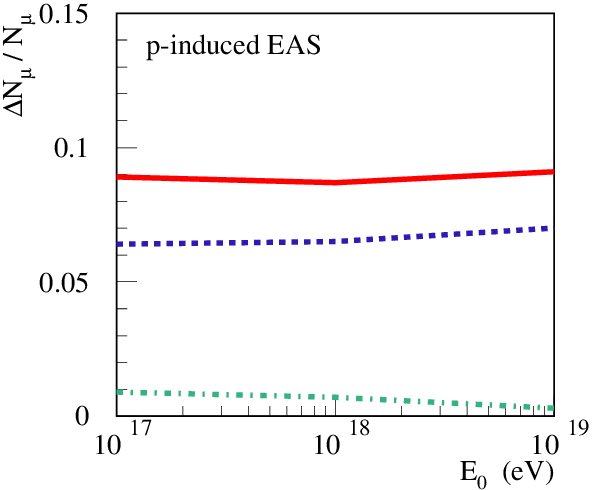}\hfill
\includegraphics[height=6cm,width=0.49\textwidth]{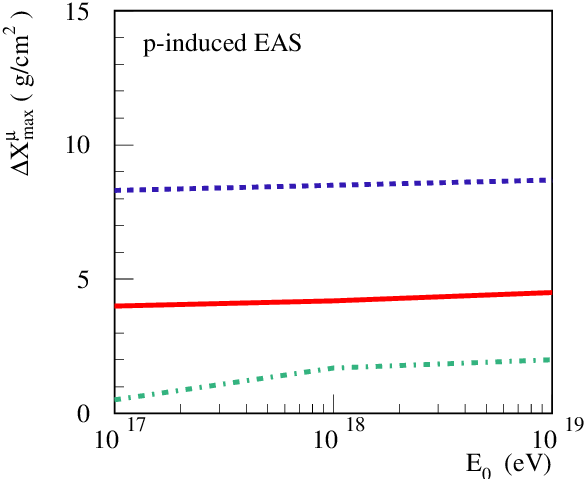}
\caption{Energy dependence of the relative change of the muon number $N_{\mu}$ at sea level (left) 
and of the modification of the maximal muon production depth  $X^{\mu}_{\max}$ (right),
both for $E_{\mu}>1$ GeV, for proton-initiated EAS,
 with respect  to the corresponding predictions of the default QGSJET-III model, 
  for the considered modifications of the model: 
 enhancement of (anti)nucleon, kaon, and $\rho$ meson production --
 solid, dashed, and dash-dotted lines, respectively.}
\label{fig:modif}       
\end{figure}

 Regarding the impact on EAS muon content, 
  the considered enhancements of kaon and (anti)nuc\-leon production give rise
  up to $\sim 10$\% increase of the  predicted  $N_{\mu}$, compared to the 
  original results of QGSJET-III, as one can see in Fig.\  \ref{fig:modif} (left). 
On the other hand, the considered increase of 
   $\rho$ meson yield appears to have only a minor effect: 
    $\Delta N_{\mu}/N_{\mu}< 1$\%, which is a direct consequence of the
    isospin symmetry.\footnote{While the isospin symmetry is not  exact  
    for strong interactions, it holds to a very good accuracy thanks to 
    the small mass difference between  $u$ and $d$ quarks.}
    Indeed, since such an enhancement of  $\rho$ meson production does not
  modify  the energy partition between 
 nuclear and e/m subcascades initiated by the, respectively, charged and
 neutral pions resulting from decays of   $\rho$ mesons, we are essentially
 back to the picture discussed in Section \ref{kinem.sec}. Then, since
 this enhancement  of  $\rho$ meson yield mainly affects central
  (in c.m.\ frame) production characterized by   small energy fractions
   $x_E$ of the mesons, at sufficiently high energies, it is of minor
 importance for  $N_{\mu}$ predictions, as demonstrated in  Section \ref{kinem.sec}.
 
 The effect of the considered modifications of particle production on the
 calculated maximal muon production depth $X^{\mu}_{\max}$ is shown in 
  Fig.\  \ref{fig:modif} (right). Enhancing the yields of secondary kaons and 
  (anti)nucleons, one shifts the energy balance between hadronic and e/m 
  subcascades in favor of the former, thereby slowing down the energy ``leak''
  from the hadronic component of extensive air showers and elongating the
  nuclear cascade, as discussed in some detail in \cite{ost16}. The corresponding
  values of $X^{\mu}_{\max}$ exceed the ones of the original QGSJET-III by up
  to $\sim 10$ g/cm$^2$. On the other hand, 
 the considered increase of $\rho$ meson production has a much weaker impact
 on  $X^{\mu}_{\max}$.
 
 \subsection{Changing the energy-dependence of the pion exchange process\label{pion-exch.sec}}
 The very last remaining option regarding a potentially strong
  enhancement of EAS muon
 content is related to pion exchange process in pion-air collisions, i.e.,
 to the so-called Reggeon-Reggeon-Pomeron ($\mathbb{RRP}$) contribution, with
 $\mathbb{R}=\pi$, to secondary hadron production. As discussed in \cite{ost13},
 assuming a dominance of the pion exchange in the  $\mathbb{RRP}$ configuration,
due to the small pion mass, compared to other  Reggeons, one obtains
 $\simeq 20$\%
enhancement of  $N_{\mu}$, relative to the case when this contribution is
neglected. This is because such a process changes the energy balance between 
secondary charged and neutral pions in favor of the former. Indeed, the dominant
decay channels $\rho^{\pm} \rightarrow \pi^{\pm}\pi^0$ and 
$\rho^{0} \rightarrow \pi^{+}\pi^-$ give rise to 
\begin{equation}
\langle x_{\pi^{\pm}}^{\mathbb{RRP}}\rangle {\rm :}
\langle x_{\pi^{0}}^{\mathbb{RRP}}\rangle
= 3{\rm :}1\, ,
\label{eq:x-part}
\end{equation}
 for the products of forward $\rho$ meson decays,
 in contrast to $\langle x_{\pi^{\pm}}\rangle {\rm :}
\langle x_{\pi^{0}}\rangle = 2{\rm :}1$ for the usual (``central'')
 $\rho$ meson production discussed in Section \ref{model-tune.sec}.

However, the energy-dependence of that contribution
 depends crucially on the treatment of absorptive effects defining the 
 probability for the corresponding ``rapidity gap survival'', i.e.,
 that the (small) rapidity gap between the leading hadron (here,  $\rho$ meson)
  and  other secondary particles produced by an interaction of the virtual
    pion with the target nucleus, is not filled
 by particle production due to additional multiple scattering processes \cite{ost13,ost23,kop15,kmr17,ost21}. 
 
 Here we are going to consider an extreme case: neglecting such absorptive
  corrections and assuming that  the
pion exchange process constitutes a fixed fraction $w_{\pi}^{\rm lim}$
 of the inelastic cross section. I.e., for any inelastic pion-air collision,
 we consider with the probability $w_{\pi}^{\rm lim}$ a production of a charged
 or neutral $\rho$ meson, with equal probabilities (e.g.\ \cite{ara91}), sampling its 
 light cone momentum fraction
 $x$ and transverse momentum $p_t$ 
  according to the corresponding probability $f_{\pi -{\rm air}}(x,p_t^2)$ 
  for reggeized pion exchange in the Born approximation (e.g.\ \cite{kai06}),
while treating the rest of hadron production for the event as an 
 interaction of the  virtual pion with air nucleus. Here
 \begin{equation}
 f_{\pi -{\rm air}}(x,p_t^2)\propto \frac{xt}{(t-m_{\pi}^2)^2}\,F^2(t)
(1-x)^{1-2\alpha_{\pi}(t)}\,  \sigma_{\pi -{\rm air}}^{\rm tot}((1-x)s)\, ,
 \label{eq:bare-pi}
 \end{equation}
with the squared momentum transfer $t$ being related to $p_t$  as
\begin{equation}
-t=p_{\perp}^2/x+(1-x)(m_{\rho}^2/x-m_{\pi}^2)\,.  \label{t-pip}
\end{equation}
$\alpha_{\pi}(t)=\alpha_{\pi}'\,(t-m_{\pi}^2)$ is
 the pion Regge trajectory with the slope $\alpha_{\pi}'\simeq 0.9$ GeV$^{-2}$,
 $\sigma^{\rm tot}_{\pi -{\rm air}}$ is the total pion-air cross
section,\footnote{Since the scattering proceeds close to the pion pole,
 it is characterized by relatively small momentum transfer squared $t$,
  i.e., we deal with quasi-real pions. Therefore, for 
   $\sigma^{\rm tot}_{\pi -{\rm air}}$ and for its  decomposition into 
   inelastic and elastic parts, Eq.\ (\ref{eq:sigpi}), we use the
 results of QGSJET-III, corresponding to the usual pion-nucleus interaction.}
 $F(t)$ is the pion emission form factor,\footnote{Since the precise form of this
 form factor is of minor importance for the current study, we use the one
 for $pp\rightarrow nX$ reaction,  $F(t)=\exp (R_{\pi}^2t/2)$, with 
$R_{\pi}^2\simeq 0.3$ GeV$^{-2}$ \cite{kop15}.}
  $m_{\pi}$ and   $m_{\rho}$ are the  pion and $\rho$ meson  masses,
  respectively, and  $s$ is the  c.m.\ energy for the interaction.

\begin{figure}[htb]
\centering
\includegraphics[height=6cm,width=0.5\textwidth]{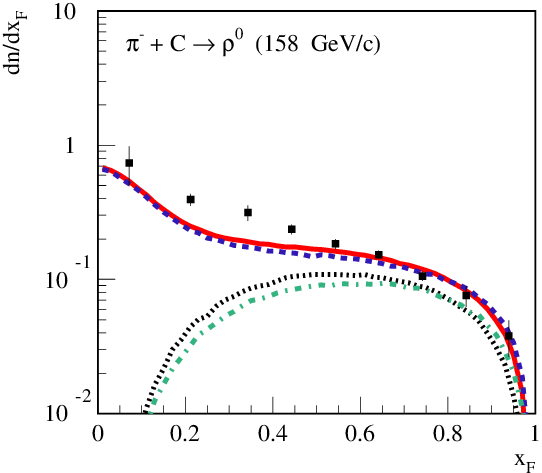}
\caption{Feynman $x$ distribution (in c.m.\ frame) of $\rho^0$ mesons produced
 in $\pi^-$C collisions at 158 GeV/c,  calculated using the default QGSJET-III
 model (solid line) or applying the alternative
treatment for the pion exchange process, as discussed in the text (dashed line).
The corresponding partial contributions of  the pion exchange process are 
shown by dotted and dash-dotted lines, respectively.}
\label{fig:rho158}       
\end{figure}

\begin{figure}[htb]
\centering
\includegraphics[height=6cm,width=0.5\textwidth]{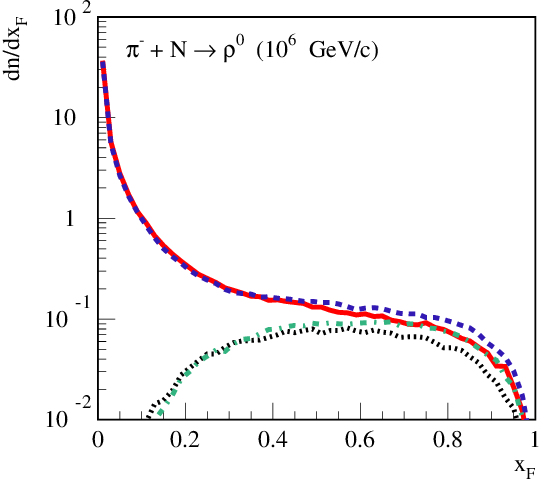}
\caption{Same as in Fig.\ \ref{fig:rho158},
 in $\pi^-$N collisions at 1 PeV.}
\label{fig:rho1pev}       
\end{figure}

\begin{figure}[htb]
\centering
\includegraphics[height=6cm,width=0.49\textwidth]{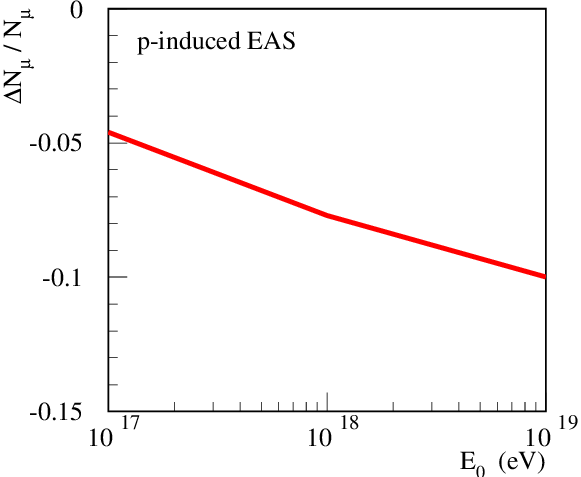}\hfill
\includegraphics[height=6cm,width=0.49\textwidth]{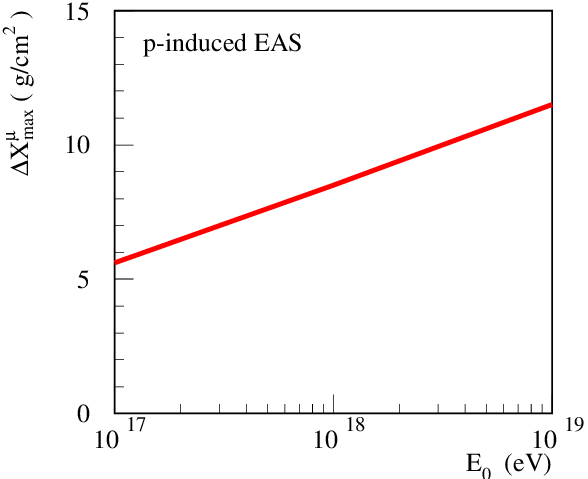}
\caption{Energy dependence of the relative change of the muon number $N_{\mu}$ at sea level (left) 
and of the modification of the maximal muon production depth  $X^{\mu}_{\max}$ (right),
both for $E_{\mu}>1$ GeV, for proton-initiated EAS,
 with respect  to the corresponding predictions of the default QGSJET-III model, 
 for  the alternative treatment of the pion exchange process,  discussed in the text.}
\label{fig:modifpi}       
\end{figure}

 Choosing $w_{\pi}^{\rm lim}$=0.11, based on the
 data of the NA61 experiment \cite{adu17}, see Fig.\  \ref{fig:rho158}, the resulting 
spectrum of $\rho^0$
 mesons and the partial contribution of the pion exchange process for 
 pion-nitrogen collisions at 1 PeV  are shown in   Fig.\  \ref{fig:rho1pev}, 
 in comparison to the
 corresponding predictions of QGSJET-III. In turn, the obtained 
 energy-dependence of the relative change of $N_{\mu}$,
  corresponding to this alternative treatment, 
is plotted in Fig.\  \ref{fig:modifpi} (left).

\begin{figure}[htb]
\centering
\includegraphics[height=4.cm,width=0.25\textwidth]{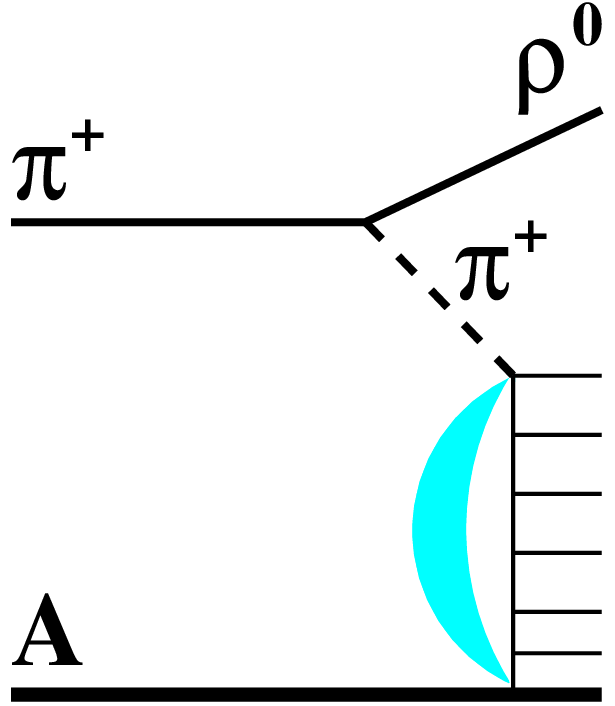}\hspace{2cm}
\includegraphics[height=4.cm,width=0.25\textwidth]{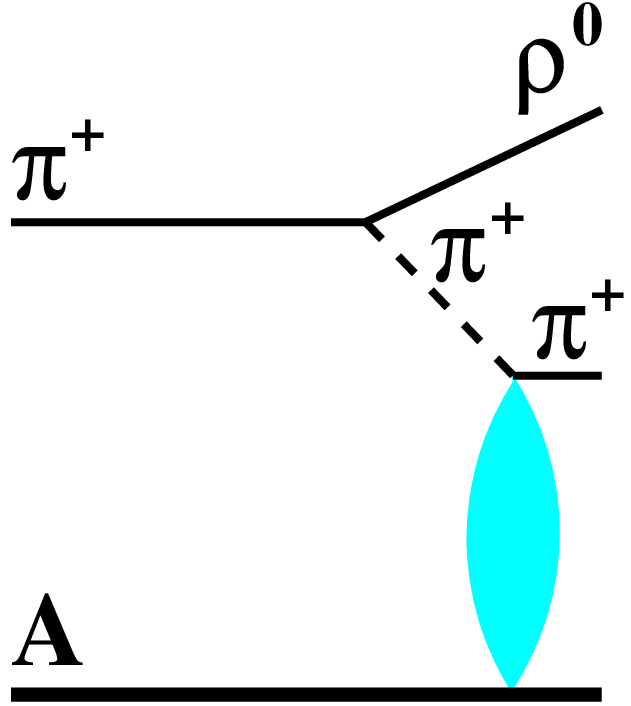}
\caption{Schematic view of the two contributions to the pion exchange process:
for inelastic scattering of the virtual pion (left),  the leading $\rho$ meson
is accompanied by multiple hadron production;  elastic scattering  (right)
gives rise to only one secondary pion, in addition to the 
 $\rho$ meson. The light shaded  croissant and ellipsis  correspond to 
 contributions of elastic rescattering processes.}
\label{fig:pitot}       
\end{figure}

Somewhat surprisingly, such a treatment results in a lower muon number,
 compared to the  predictions of the 
default QGSJET-III model. To understand this nontrivial result, let us remind 
ourselves that a considerable part of the cross section
 $\sigma_{\pi-{\rm air}}^{\rm tot}$ in Eq.\  (\ref{eq:bare-pi}) corresponds to
 elastic scattering (see Fig.\ \ref{fig:pitot}), 
 \begin{equation} 
 \sigma_{\pi-{\rm air}}^{\rm tot}=\sigma_{\pi-{\rm air}}^{\rm el}
 +\sigma_{\pi-{\rm air}}^{\rm inel}\,,
 \label{eq:sigpi}
 \end{equation} 
 with the corresponding fraction
 $\sigma_{\pi-{\rm air}}^{\rm el}/\sigma_{\pi-{\rm air}}^{\rm tot}$ being $\sim 30$\%
 at fixed target energies and slowly rising with $E_0$, approaching the ``black
 disk'' limit 
 ($\sigma_{\pi-{\rm air}}^{\rm el}/\sigma_{\pi-{\rm air}}^{\rm tot}\rightarrow 0.5$)
 in the $E_0\rightarrow \infty$ limit. For that elastic scattering contribution,
 the final state consists of  a $\rho$ meson and a pion only: with the virtual
 pion being put on-shell by the scattering process, see  Fig.\ \ref{fig:pitot} (right).
  It is such a scarce hadron  production which gives rise to the
   obtained decrease of  $N_{\mu}$.

In contrast, in the treatment of the  QGSJET-III model, the rise of absorptive
corrections ``pushes'' the pion exchange process towards larger and 
larger impact
parameters, causing a (slow) decrease of its relative contribution \cite{ost23,ost21}.
At the same time, at large impact parameters, the above-discussed contribution of
elastic scattering of the virtual pion is strongly suppressed and can be 
neglected \cite{ost21}.

Regarding the maximal muon production depth, in addition to a scarce hadron production
in case of elastic scattering of the virtual pion, one has a shift of the  
 energy balance between hadronic and e/m 
  subcascades in favor of the former [cf.\ Eq.\ (\ref{eq:x-part})],
   both effects contributing to an elongation
  of the nuclear cascade profile. The resulting  $X^{\mu}_{\max}$ values exceed the 
  ones of the original QGSJET-III by up to $\sim 10$ g/cm$^2$, see  Fig.\  \ref{fig:modifpi} (right).

\section{Conclusions\label{summary.sec}}
In this work, we investigated the possibility to increase the predicted EAS muon content,
within the framework of the QGSJET-III model, restricting ourselves to the standard
physics picture. We started with  specifying the 
 kinematic range for secondary hadron production, which is of relevance 
 for  model predictions regarding the  muon number $N_{\mu}$ and the 
 maximal muon  production depth $X^{\mu}_{\max}$, and demonstrated that these
 quantities are governed by forward pion production in hadron-air collisions:
  with the corresponding pion energy fractions, taken from their parent hadrons,
   ranging between few
   percent and few tens of percent. Further, we investigated the impact
 on the predicted $N_{\mu}$ and  $X^{\mu}_{\max}$
 of various modifications of the  treatment of hadronic interactions,
in particular the model calibration to accelerator data, the amount of 
the ``glue'' in the pion, and the energy dependence of the pion exchange process. 

The strongest $N_{\mu}$ enhancement reaching $\sim 10$\% level has been obtained 
when we increased the yields of secondary kaons and (anti)nucleons in the
model, adjusting its predictions to NA61 data on pion-carbon interactions
at fixed target energies. However, such  modifications should be considered
as  extreme ones, since they lead to a serious tension with results of other
experiments, regarding the yields of such hadrons both in pion-proton and,
especially, in proton-proton collisions.

As for  the gluon content of the pion, even considering extreme changes of
the momentum partition between valence quarks and gluons, we obtained 
rather insignificant modifications of 
the predicted $N_{\mu}$ and  $X^{\mu}_{\max}$.
 This is because such changes impact
 pion-air collisions at the highest energies only, in the atmospheric
nuclear cascade, while being of minor importance for the bulk of such
interactions at lower energies, at later stages of the cascade development.

Regarding the  pion exchange process in pion-air collisions, 
the (slow) decrease with energy  of the corresponding probability can only
be tamed if one assumes that absorptive corrections to the process are
rather weak. However, in such a case, a significant part of the cross
section for the virtual pion interaction with the target nucleus corresponds
to elastic scattering. The corresponding very scarce hadron production
results in a decrease of the  predicted $N_{\mu}$.

Regarding $X^{\mu}_{\max}$, for some of the considered
 modifications of particle production, we obtained up to $\sim 10$  g/cm$^2$
 changes of  the  maximal muon  production depth, 
 typically yielding larger values of   $X^{\mu}_{\max}$.
 This aggravates the tension with observations of the Pierre Auger Observatory.

Overall, we see no possibility to increase the predicted EAS muon content
by more than 10\%, without entering in a serious contradiction with relevant
accelerator data, while staying within  the standard physics picture.

\subsection*{Acknowledgments}
 The work of S.O.\ was supported by  Deutsche Forschungsgemeinschaft 
(project number 465275045).  G.S.\ acknowledges
support by the Bundesministerium f\"ur Bildung
und Forschung, under grants 05A20GU2 and 05A23GU3.

\end{document}